# CLOSE NEIGHBORS OF MARKARIAN GALAXIES. I. OPTICAL DATABASE

*T.A. Nazaryan[1], A.R. Petrosian[1], B.J. Mclean[2]*

*The catalogue of close neighbors of Markarian galaxies located inside of circles with radii 60 kpc from the centers of Markarian objects is presented, which combines extensive new measurements of their optical parameters with a literature and database search. The measurements were made using images extracted from the Digitized Sky Survey (DSS) $J_{pg}$ (blue), $F_{pg}$ (red) and $I_{pg}$ (near-infrared) band photographic plates. We provide names, accurate coordinates, redshifts, morphological types, blue, red and near-infrared apparent magnitudes, apparent blue major diameters, axial ratios, as well as position angles for the neighbor galaxies. We also include their 2MASS infrared magnitudes. The total number of Markarian galaxies in the database is 274 and number of their neighbors is 359. The physical parameters of the systems of Markarian galaxies and their neighbors are determined and presented.*

## 1. Introduction.

Studying the environments of galaxies is an important way to understand their origin and evolution. The properties of galaxies such as morphology, luminosity, stellar population, star formation rate etc. show correlations both with their large-scale environments (e.g. [1] for review) as well presence of near companions (e.g.[2]). The activity events in galaxies such as AGNs and the intensive star formation that can influence subsequent evolution of galaxies vastly are proposed sometimes to be connected with the tidal and merging influences on their host galaxy by its neighbors, as one can conclude from observations [3-5]. Several theoretical models were proposed to describe the effects of environments on AGNs and nuclear starburst activities such as falling of interstellar gas on nucleus because of the gravitational perturbations, gas turbulence and fragmentation into clouds [6,7]. Although theoretical models succeeded in explanation of main features of tidally triggered activity and star formation in the galaxies, their results may strongly depend on parameters of simulations [8], and they cannot explain all the features and details of the relation between different forms of galaxies' activity and their environments, particularly the relation between non-nuclear starbursts or host galaxy star formation intensity and interaction, see [9] for review. So the reasonable conclusion would be that the activity of galaxies and/or their star formation may depend also on some preexisting conditions or on some combinations of galaxies' internal and external interaction properties only to some extent, see e.g. [7,9].

The objects from the First Byurakan Survey, which are more known as Markarian galaxies are well studied and have mostly AGN and/or enhanced star-forming properties. Most comprehensive datasets including all Markarian objects and their different data are contained in three catalogs [10-12]. Close environmental properties of Markarian galaxies and their connection with galaxies' inner parameters were studied in details in number of papers, first of which was the article [13] on Markarian-Markarian galaxies pairs selected from the first five lists of Markarian survey. Later Karachentsev [14], studying 65 pairs of galaxies in which at least one component is Markarian object, it has been concluded that Markarians paired with their neighbors form systems different by their dynamical parameters of those of paired normal galaxies. In [15] pairing properties of 516 non-Seyfert Markarian galaxies have been studied and it has been found out that these galaxies are exceptionally likely to occur in tight pairs, but the authors were unable to identify any clear difference between global characteristics of paired and non-paired Markarian galaxies. In two samples in [16,17] and following detailed studies (e.g. [18]), the class of double and multiple nuclei Markarian galaxies was introduced. Later, large numbers of the studies have been addressed to these objects. For example, in [19] the Infrared Astronomical Satellite (IRAS) data for 187 Markarian galaxies with multiple optical nuclei or extreme morphological peculiarities have been investigated, and by means of statistical comparison with the sample galaxies it have been found out that Markarian galaxies have significantly higher median dust temperature. The authors showed that enhancement of far infrared luminosity induced by galaxy collisions depends on assumed nature of precursor galaxies strongly and that color temperature correlates with projected nuclear separation significantly. Many double and multiple nuclei Markarian galaxies were studied individually. For example, Markarian 266, which was first studied thoroughly in [18] in 1980, currently may account more than one dozen detailed investigations in different wavelengths (e.g. [20-22]). In recent years, attention to the problem concerning the role of close interactions as a possible triggering mechanism of the activity of AGN and starburst galaxies has increased dramatically (e.g. [23]). Markarian galaxies and their close neighbors are often included in the samples of mentioned studies (e.g. [24]), but, similar to [15], there is no study based on the sample of Markarian galaxies and their neighbors only.

The large amount of new multi-wavelength observational data collected mostly in different databases (e.g. the Sloan Digital Sky Survey(SDSS), Two Micron All Sky Survey (2MASS), ROSAT, the NRAO VLA Sky Survey (NVSS), the HI Parkes All Sky Survey (HIPASS)) during the last two decades, as well as our new homogeneous measurement, made possible to collect large data-sets containing more complete information about the neighbor objects of Markarian galaxies as well as their own properties, which can be useful for systematical study of these systems and particularly answer the following questions:

1). What is the true connection between galaxy-galaxy interactions, nuclear activity and star formation?

2). Is there any correlation between integral parameters of Markarian galaxies and their neighbors?

3). How the activity levels of Markarian galaxies and their star forming properties are correlated with the similar properties of their close neighbors?

4). Is the linear distance between Markarian galaxy and its neighbor the only or one of the crucial parameters that determines correlation between physical properties of two galaxies?

This work is the first in the series of the studies for comparative statistical research of properties of galaxies that are close neighbors of the galaxies with different levels of nuclear activity and star-forming properties. These studies aim to find clues toward disclosure of the role and contribution of galaxies' interactions and merging processes in their formation and evolution, particularly understanding their activity and star forming properties. In general, samples of the objects for mentioned study include Markarian and Second Byurakan Survey (SBS) galaxies, as well as so called passive galaxies and their close neighbors. This paper presents optical data-set for Markarian galaxies and their close neighbors. The method of construction of the sample of Markarian galaxies' close neighbors is presented and its efficiency and possible selection effects are evaluated in section 2 of this paper. The observational material used and generation of the database are described in section 3. The database itself and the related information are described in section 4. In section 5 we summarize obtained data.

**2. The method of search.**

We sought galaxies close to Markarian galaxies in position-redshift space as the most probable candidates of physical neighbors. The projected separation between possible physical neighbors may vary between a few and several hundreds of kpcs. It has been shown that neighbors that may have significant tidal effects on each other and able to trigger star formation have threshold distance of about 150 kpc, and the strongest induced star formation events are in galaxies which have characteristic separation threshold ~40 kpc independently of environment [25], [26]. Projected separation corresponding to 50 kpc was used in paper [12] to find the number of possible neighbors of Markarian galaxies. In this study, physical neighbor(s) for Markarian galaxies are selected within angular separation corresponding to 60 kpc projected distance (in this study for Hubble constant the value of $H_0 = 73\ km\, s^{-1} Mpc^{-1}$ is considered). For separation criteria we considered 60 kpc, instead of 50 kpc [12], in order to better meet above mentioned threshold for inducing strong star formation and also to be sure that we include all the neighbors from [12] taking into account possible coordinates uncertainties and large angular sizes of the galaxies. In different studies, to distinguish physical connection between galaxies, different values for their radial velocities differences have been accepted. Accepted values are in broad range and can reach up to $1000\ km\, s^{-1}$ (e.g. [4]). In this study, in order to select physical neighbor(s) for Markarian galaxies we took the criteria $\Delta v_r = 800\ km\, s^{-1}$ for difference of radial velocities of neighbors with respect to Markarian galaxy. This accepted value is about twice larger than difference of radial velocities which is the characterizing threshold for inducing significant star formation [25], but it was selected to escape data pollution and not "to lose", in general, those physical systems which can have higher velocity dispersions but may contain members with significant mutual tidal influences on each other (e.g. [25]). Using this criteria, we counted neighbors around all Markarian galaxies that have redshift $z \geq 0.005$ to be sure that random projection effects and relative error because of low redshift are not significant.

The number of neighbors around each Markarian galaxy is counted in paper [12] using the following criteria: 1) redshift of Markarian galaxy is larger than 0.005, 2) projected distance of neighbor lies within 50 kpc, and 3) angular sizes of galaxies differ from those of the Markarian galaxy by no more than factor of 2. In [12] 1424 Markarian galaxies have defined redshift larger than 0.005. It was found that 448 galaxies among them have 640 neighbors total in [12]. In the current study by using the same redshift value limitation (redshift of Markarian galaxy is larger than 0.005) and by using the criteria $\Delta v_r = 800\, km\, s^{-1}$ for difference of radial velocities of neighbors with respect to Markarian galaxy we have found that 274 Markarian galaxies have 359 neighbors within 60 kpc projected distance. Comparatively, within 50 kpc distance from Markarian galaxy [12], 230 Markarian galaxies have 312 neighbors. By comparing the objects found in the current study and in [12], we found that there are 395 objects that were identified as neighbors in [12], but we did not find them, mainly because of having significantly different redshifts from those of Markarians or because neighbor(s) redshift(s) is(are) not measured. Also there are 67 objects that are identified as neighbors in our study but were not found in [12] mainly because of having too small or too large angular size (less or more than 2 times) compared to that of the Markarian galaxy. They are included in our database. 152 Markarian galaxies turned out to have the same amount of neighbors.

We analyzed our sample of neighbors for several possible selection effects. We checked two possible selection effects: 1). Is our sample of neighbors biased against those Markarain galaxies that are not located within the part of sky covered by SDSS? Simple comparison between number of neighbors and Markarian galaxies with and without SDSS images shows that it can be neglected in comparison with the random statistical distributions. 2). Is our sample of neighbors biased against some close neighbors of distant Markarian galaxies? Analysis of the sample showed that this effect also can be neglected.

Incompleteness of the sample of neighbors was estimated by comparing results of our search for neighbors and that done in [12] following the method presented in [25] for comparison of spectroscopic sample of possible neighbors with the photometric one. We selected subsample of Markarian galaxies and analyzed those Markarian galaxies that have different amount of neighbors found in this study and in [12] and inspected their 50 kpc surroundings visually to understand why numbers of discovered neighbors are different. We calculated the amount of galaxies that were not found by us, but those were found in [12] so that their redshift is unknown. Those galaxies are potential neighbors of Markarian galaxies with some probability. We also calculated the number of galaxies that were not found by us, but that were found in [12] so that their redshift is known and is not within $\Delta v_r = 800\, km\, s^{-1}$ interval compared to the redshift of the corresponding Markarian galaxy. Comparison of this number with the amount of neighbors (of Markarian galaxies of selected subsample) with known redshifts that lie within $\Delta v_r = 800\, km\, s^{-1}$ gave us an opportunity to estimate probability for the galaxy within 50 kpc neighborhood to have a close redshift and to be a real physical neighbor. We estimated the incompleteness of our sample of neighbors as about 15-18 per cent by using this method.

## 3. The observational material and the sample.

The search of neighbors around Markarian galaxies was carried out using above mentioned two, linear separation and radial velocity difference criteria. Radial velocities for the potential neighbors of Markarian galaxies within 60 kpc circles were checked using two data sources: NASA/IPAC Extragalactic Database (NED) and Sloan Digital Sky Survey Data Release 8 (SDSS DR8). Almost all redshifts for neighbors are collected from the NED, which also includes data from SDSS DR 1 to 6. Redshifts are determined according to SDSS DR8 for remaining number of neighbors. In [12] it was found out that almost 70% of Markarian galaxies are in the fields of SDSS DR5 images, this number was increased to 87% due to SDSS DR8. 12 faint galaxies in our database are not included in NED (or do not have defined redshift in NED) and were identified only using their redshifts and images in SDSS DR8.

We used DSSII multi-color images to check positions of neighbors obtained from NED and SDSS DR8, to make primary morphological classification, to measure angular sizes and position angles and to find apparent magnitudes in three J, F and I wavebands. DSSII is based on POSS-II (north) and UKSTU (south) surveys. The Palomar Sky Survey (POSS-II, [27]) covers Northern sky above equator, it has been done in three wavebands: blue J (IIIa-J emulsion and GC395 filter: $\lambda_{eff} \sim 4800 \mathring{A}$ ), red F ( IIIa-F emulsion and RG610 filter: $\lambda_{eff} \sim 6500 \mathring{A}$ ) and near-infrared I ( INV emulsion and RG9 filter: $\lambda_{eff} \sim 8500 \mathring{A}$ ). The ESO/SERC survey [28] covers Southern hemisphere and has been done in the same wavebands. The limiting magnitude for POSS-II J band is $J_{pg} = 22.5$ (23.0 for UKSTU SERC-EJ), for POSS-II F band it is $F_{pg} = 20.8$ (22.0 for UKSTU SERC-ER), and for POSS-II I band it is $I_{pg} = 19.5$ (19.5 for UKSTU SERC-I). All the plates were scanned at the Space Telescope Science Institute (STScI) with $1.0'' px^{-1}$ resolution ( SERC-ER plates were scanned with $1.7'' px^{-1}$ ). Color images of the galaxies available in SDSS DR8, were used to check neighbors identification and verify their morphological classification done by means of DSSII multi-color images.

***3.1. Coordinates***. The coordinates of the neighbors were obtained by a two-step algorithm. First, we found neighbors via automatic search procedure and extracted their coordinates from NED or from SDSS DR8 images. Then this coordinates were carefully checked by the DSSII blue images. Positions of compact objects are typically located to better than 0.5 arcsec using either a 2D Gaussian fit or the intensity-weighted moments of the object pixels. The actual positions of the extended objects are somewhat more poorly determined, because of their more complex morphologies and difficulty of locating the image centroid of the objects. Positions were measured using peak intensity of the objects for these galaxies. In these cases the positions may be uncertain to $1''$. All coordinates are in the HST Guide Star Selection J2000.0 System.

***3.2. Redshifts***. In this database, heliocentric redshifts for neighbors were collected from NED and SDSS DR8. Redshifts of Markarian galaxies were taken from [12]. 1524 Markarian galaxies have defined redshifts in that database, and 1424 of them have redshifts more than 0.005. Neighbors in our database have radial velocities within $\Delta v_r = 800\ km\ s^{-1}$ interval compared to that of the corresponding Markarian galaxy. About 86% of neighbors lie within more narrower interval of difference of radial velocities $\Delta v_r = 300\ km\ s^{-1}$.

***3.3. Morphology.*** The current database presents complete and homogeneous morphological classification of physical neighbors of Markarian galaxies. DSSII blue plates were used as primary source for galaxies morphological classification. The uniformity of these survey plates makes them useful for examining structural features of bright (13-16 mag.) as well as relatively faint galaxies, both the central parts and outer regions. DSSII red and near-infrared plates were used as supplementary material to check morphological structure of the galaxies. For morphological classification we used not only images, but also isophotal maps of all galaxies, which were constructed to display the large dynamic range of the images and were extremely useful for classification of faint and small angular size objects. For the neighbors on SDSS frames DSSII morphological classification was verified also by SDSS color images.

All neighbors of Markarian galaxies were classified morphologically according to the modified Hubble sequence (E-S0-Sa-Sb-Sc-Sd-Sm-Im) and the extension to blue compact dwarf galaxies (BCDs). Following [12] we classified irregular galaxies with one giant HII region as Im/BCD or BCD/Im depending on what component is dominant, extended Im system or more compact, bright HII region. Close interacting systems and mergers are morphologically classified as separate class of objects. Figure1 shows comparison between our determined morphological classes for all 359 objects and 160 of them that have defined morphologies in Leon-Meudon Extragalactic Database (HYPERLEDA). There are 2 galaxies (IC1640 and NGC 2968), which were classified in HYPERLEDA as of late-type, but we classified them as of early-type galaxies. 4 galaxies (MCG +12-09-021, NGC 7464, NGC 4004B, MCG +07-33-040) were classified in HYPERLEDA as of early-type but we classified them as of late-type. In HYPERLEDA 4 galaxies (2MASX J15563687+4154131, NGC 3010. 2MASX J03195122+4132072, PGC 012450) are marked as possible spiral galaxies ("S?"), we did exact classification of these objects and classified them as of early-type.

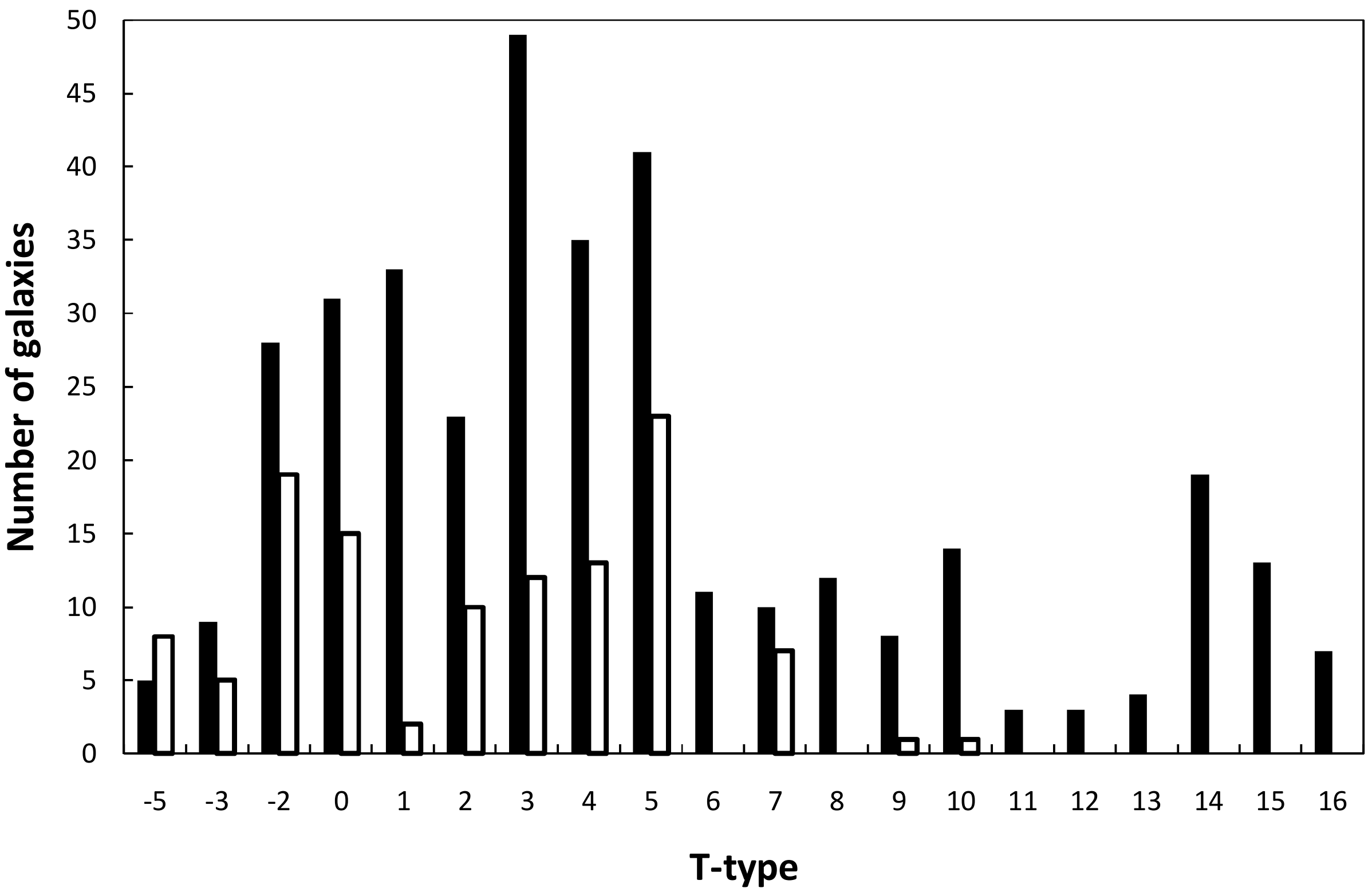

Fig. 1. Comparison of our (filled bars) and HYPERLEDA (empty bars) T-types of neighbors.

***3.4. Apparent $J_{pg}$ (blue), $F_{pg}$ (red), $I_{pg}$ (near-infrared) magnitudes***. In this database we provide our measurements of the apparent blue, red and near-infrared magnitudes of neighbors of Markarian galaxies in accurate and homogeneous manner. The magnitudes of the galaxies were measured from the POSS-II and UKSTU photographic survey plates, collected in DSSII, that are available at STScI and used for the construction of the guide Star Catalog II (GSC-II) [29]. In order to measure the magnitudes of nonstellar objects, it was necessary to use a technique to derive the nonlinear intensity to photographic density transformation so that one can integrate in intensity units. We have modified the technique developed by the APM group (e.g. [30]) to calibrate photographic plates internally using the measured stellar profiles. This is an iterative process, which as a by-product determines the photographic density – intensity ( *D–I* ) function for that plate. By adding a step to this iterative process of at least-squares fit of the derived magnitudes of the GSC-II stars in the field to their catalog magnitudes we effectively constrain the calibration to be linear with a determined zero point. One can then take the D-I function and zero point to integrate the intensity within the isophotal contours of the galaxy and convert that to a magnitude. In cases where we were unable to automatically de-blend overlapping objects in software, a manual procedure was used to mark boundary between objects in order to assign the pixel's surface brightness to the correct object. This does not eliminate measurement errors due to the saturation that can occur in the nuclei of the brighter galaxies or possible field effects where the *D-I* response varies over the plate. The blue, red and near-infrared magnitudes of galaxies were measured from the $J_{pg},\ F_{pg},\ I_{pg}$ band images of the objects in a homogeneous way at the isophote corresponding to 3 times the background *rms* noise above the background or $3\sigma$, which is approximately $25.2\ mag\ arcsec^{-2}$ [12].

***3.5. Major and minor angular diameters and axial ratios.*** The major and minor diameters of neighbors of Markarian galaxies were measured in a homogeneous way using DSSII blue images at the isophote corresponding to 3 times background *rms* noise. Surface brightness level diameters correspond to the $Bj = 25.2\ mag\ arcsec^{-2}$ limiting surface brightness diameters [12].

HYPERLEDA data was used to verify the agreement of our measured diameters with the standard D(25) diameter system. Comparison was done separately for early and late-type galaxies. We excluded objects that have significantly different diameters in HYPERLEDA mostly because of misidentifications. For early and late-type galaxies the linear correlation between our and HYPERLEDA D(25) diameters have the following forms:

early-type: $D''(25) = (0.902 \pm 0.026)D''(J) + (5.411 \pm 1.715),\ r = 0.953, N = 61,$

late-type: $D''(25) = (0.896 \pm 0.020)D''(J) + (1.944 \pm 1.190),\ r = 0.928, N = 158.$

The fact that blue diameters measured by us are larger than those of HYPERLEDA is because of the fainter surface brightness limit we used.

Using diameters we also calculated axial ratios of galaxies and again compared them with the HYPERLEDA axial ratios. The linear correlation between these axial ratios has the following form for early-type and late-type galaxies respectively:

early-type: $R(25) = (0.939 \pm 0.066)R(J) + (0.035 \pm 0.047),\ r = 0.776, N = 59,$

late-type: $R(25) = (0.877 \pm 0.040)R(J) + (0.074 \pm 0.024),\ r = 0.763, N = 155.$

Both the diameters and axial ratios of galaxies in HYPERLEDA and our measurements have no significant differences.

***3.6. Position angles.*** The position angles (PA) of the major axes of the galaxies were determined at the same $25.2\ mag\ arcsec^{-2}$ isophotal level. PA is measured from the North ($PA = 0°$) toward the East between $0°$ and $180°$. Determined by the 49 common early-type galaxies, the average absolute difference of position angles measured by us and in HYPERLEDA is $|\Delta PA| = 8.8° \pm 1.2°$. The same difference determined for 154 late-type galaxies is $|\Delta PA| = 8.4° \pm 0.6°$. PAs for the galaxies with axial ratios close to 1 were not measured.

**4. The database.**

Table 1 contains the observational data for 319 galaxies aligned in 15 columns which are described below (this table does not include parameters of multi-component Markarian galaxies, that can be found in [12]). In 32 cases Markarian galaxies form pairs with other Markarian object. For these Markarian neighbor galaxies integral parameters are copied from [12]. Only exception is $I_{pg}$ magnitudes which were measured during this study.

Column (1) – Markarian galaxy name (*MRK*) according to [12].

Column (2) – Number of individual neighbor(s) (*N*) for the given Markarian galaxy.

Column (3) – Name of the neighbor galaxy (*Neighbor)*. If it has names in different catalogs, the most recognizable of them is used.

Column (4) – Right ascension (*RA*) of the neighbor galaxy (J2000.0 equinox)

Column (5) – Declination (*Dec*) of the neighbor galaxy (J2000.0 equinox).

Column (6) – Heliocentric redshift (*z*) of the neighbor galaxy.

Column (7) – Morphological description (*Morph*) of the neighbor galaxy. The numerical coding used here for the morphological description of the galaxies is a slightly modified and simplified version of the morphological types T given in the Third Reference Catalogue of Bright Galaxies (RC3). The following codes were used: E = -5; E/S0 = -3; S0 = -2; S0/a = 0; Sa = 1; Sab = 2; Sb = 3; Sbc = 4; Sc = 5; Scd = 6; Sd = 7; Sdm = 8; Sm = 9; Im = 10; Im/BCD = 11; BCD/Im = 12; BCD = 13; Compact = 14; Interacting system or Merger = 15; HII region = 16; An existence of bar is marked by "B".

Column (8) – DSS-II apparent $J_{pg},\ F_{pg},\ I_{pg}$ band magnitudes respectively.

Column (9) – Major $J_{pg}$ band *D* diameter in arcseconds.

Column (10) – $J_{pg}$ band axial ratio *R*.

Column (11) – Position angle in $J_{pg}$ band *PA*. It is measured from north ($PA = 0°$) toward east in $[0°, 180°)$. For round galaxies with axial ratios close to 1 position angles are not determined (ND).

Column (12) – 2MASS [31] *J, H* and *K* band magnitudes respectively if available.

Column (13) – Projected distance *(d)* from the Markarian galaxy in arcseconds.

Column (14) – Absolute value of difference of radial velocity ($\Delta v_r$) with respect to the Markarian galaxy in $km\ s^{-1}$ units.

Column (15) – Decimal logarithm of estimated mass to blue band luminosity ratio *(log(M/L))* in solar mass and luminosity units for the system of Markarian galaxy and its neighbor. This parameter is calculated assuming that the Markarian galaxy and its neighbor are in circular motion around common center. The following formula is used to estimate the total mass of the system:

$$M = \frac{32}{3\pi}\frac{(\Delta v_r)^2 d_{proj}}{G}$$

Where $\Delta v_r$ is the difference of radial velocities, $d_{proj}$ is the projected separation between galaxies and $\frac{32}{3\pi}$ is the projection factor that arises from the random orientations of pairs. Virgocentric redshifts were used instead of the heliocentric redshifts while calculating the

*Table 1.*

PART OF DATABASE OF NEIGHBORS OF MARKARIAN GALAXIES

| *MRK* | *N* | *Neighbor* | *RA* | *Dec* | *z* | *Morph* | $J_{pg}$ | $F_{pg}$ | $I_{pg}$ | *D* | *R* | *PA* | *J* | *H* | *K* | *d* | $\Delta v_r$ | *Log(M/L)* |
|---|---|---|---|---|---|---|---|---|---|---|---|---|---|---|---|---|---|---|
| (1) | (2) | (3) | (4) | (5) | (6) | (7) | (8) | | | (9) | (10) | (11) | (12) | | | (13) | (14) | (15) |
| 1 | 1 | MRK 976 | 01 16 12.46 | + 33 03 51.5 | 0.0163 | 5 | 15.8 | 14.4 | 13.6 | 43.9 | 0.79 | 44 | 12.5 | 11.9 | 11.6 | 111.0 | 100 | 1.62 |
| 2 | 1 | KUG 0152+366 | 01 55 01.71 | + 36 55 10.4 | 0.0188 | 4 B | 15.8 | 15.0 | 14.4 | 62.0 | 0.54 | 64 | 13.4 | 12.7 | 12.4 | 94.6 | 0 | ND |
| 15 | 1 | UGC 04451 | 08 34 33.29 | + 75 09 12.0 | 0.0205 | 2 | 14.8 | 14.0 | 13.6 | 56.0 | 0.27 | 124 | 12.2 | 11.4 | 11.1 | 64.2 | 357 | 2.18 |
| 18 | 1 | UGC 04727 | 09 01 43.57 | + 60 09 26.2 | 0.0108 | 5 | 16.2 | 15.6 | 15.4 | 82.5 | 0.17 | 81 | | | | 112.1 | 89 | 1.41 |
| 30 | 1 | MRK 31 | 10 19 42.86 | + 57 25 25.0 | 0.0260 | 2 B | 15.3 | 14.5 | 13.8 | 64.3 | 0.68 | 175 | 12.6 | 11.9 | 11.7 | 41.0 | 96 | 1.09 |
| 31 | 1 | MRK 30 | 10 19 38.31 | + 57 25 07.0 | 0.0263 | 1 | 16.6 | 15.9 | 15.5 | 29.6 | 0.52 | 85 | 15.0 | 14.3 | 13.9 | 41.0 | 96 | 1.09 |
| 37 | 1 | KUG 1113+290B | 11 16 33.89 | + 28 46 07.1 | 0.0235 | 4 | 15.4 | 15.2 | 15.0 | 23.5 | 0.98 | 23 | 14.0 | 13.4 | 13.3 | 34.2 | 44 | 0.40 |
| 38 | 1 | MRK 39 | 11 18 20.53 | + 53 45 12.4 | 0.0361 | 3 B | 16.4 | 15.5 | 15.0 | 22.4 | 1.00 | ND | 14.9 | 14.3 | 14.2 | 32.6 | 10 | -1.08 |
| 39 | 1 | MRK 38 | 11 18 17.15 | + 53 44 59.6 | 0.0361 | 4 B | 15.7 | 14.2 | 14.0 | 38.8 | 0.47 | 78 | 13.4 | 12.7 | 12.4 | 32.6 | 10 | -1.08 |
| 40 | 1 | SDSS J112535.23+5423 | 11 25 35.23 | + 54 23 14.3 | 0.0208 | 6 | 16.0 | 15.5 | 15.3 | 58.0 | 0.17 | 151 | 15.0 | 14.7 | 14.2 | 19.0 | 92 | 0.51 |
| 41 | 1 | NGC 3759 | 11 36 53.99 | + 54 49 22.1 | 0.0186 | -2 | 13.8 | 13.1 | 12.7 | 60.9 | 1.00 | ND | 11.4 | 10.6 | 10.4 | 131.0 | 231 | 1.82 |
| 43 | 1 | KUG 1200+397A | 12 02 49.80 | + 39 26 01.9 | 0.0205 | 3 | 14.7 | 14.2 | 14.0 | 38.3 | 0.75 | 46 | 12.9 | 12.3 | 12.0 | 125.0 | 8 | -0.77 |
| 56 | 1 | SDSS J125831.89+2715 | 12 58 31.89 | + 27 15 08.5 | 0.0253 | 10 | 19.7 | 19 | 18.7 | 8.6 | 0.66 | 14 | | | | 63.8 | 217 | 2.45 |

distance between Markarian galaxy and its neighbor and their luminosities. Also, in order to calculate the luminosities, magnitudes of galaxies were corrected for Galactic foreground [32] and galaxy internal [33] extinction.

The whole database is provided separately as an online catalog. A part of the database is presented in this article as an example.

## 5. Summary.

We compiled a homogeneous database of physical neighbors of Markarian galaxies discovered within 60 kpc projected distance and $800\,km\,s^{-1}$ radial velocity intervals. The database includes 359 neighbor galaxies for 274 Markarian objects with measured exact coordinates, $J_{pg}$, $F_{pg}$ and $I_{pg}$ apparent magnitudes, angular sizes, inclination and position angles, determined morphologies, collected redshifts and 2MASS *J,H,K* magnitudes for about 60% of the neighbors. Database includes also calculated mass to luminosity ratios for the systems of Markarian galaxies and their neighbors.

All parameters of neighbor galaxies were measured in the same way as it was done for Markarian galaxies in [12]. Combined information from both databases can be used to investigate the problems stressed in the introduction of this article. These studies are in progress and will be reported in forthcoming papers.

This research has made use of NASA/IPAC extragalactic database (NED), which is operated by the Jet Propulsion Laboratory, California Institute of Technology, under contract with the National Aeronautics and Space Administration and HYPERLEDA (Leon-Meudon Extragalactic Database). This publication makes use of data products from the Two Micron All Sky Survey, which is a joint project of the University of Massachusetts and the Infrared Processing and Analysis Center/California Institute of Technology, funded by the National Aeronautics and Space Administration and the National Science Foundation. The Digitized Sky Survey was produced at the Space Telescope Science Institute under US Government grant NAG W-2166. The images of this survey are based on photographic data obtained using the Oschin Schmidt Telescope on Palomar Mountain and the UK Schmidt Telescope. The plates were processed into the present digital form with the permission of these institutions. The Second Palomar Observatory Sky Survey (POSS-II) was made by the California Institute of Technology with funds from the National Science Foundation, the National Aeronautics and Space Administration, the National Geographic Society, the Sloan Foundation, the Samuel Oschin Foundation, and the Eastman Kodak Corporation. The California Institute of Technology and Palomar Observatory operate the Oschin Schmidt Telescope. Funding for the creation and distribution of the SDSS Archive has been provided by the Alfred P. Sloan Foundation, the Participating Institutions, the National Aeronautics and Space Administration, the National Science Foundation, the U.S. Department of Energy, the Japanese Monbukagakusho, and the Max Planck Society.

[1] Byurakan Astrophysical Observatory,0213 Byurakan, Aragatsotn Province, Armenia, email: nazaryan@bao.sci.am

[2] Space Telescope Science Institute, 3700 San Martin Drive, Baltimore, MD 21218, USA